\documentclass[10pt,conference]{IEEEtran}
\IEEEoverridecommandlockouts

\usepackage{cite}
\usepackage{amsmath,amssymb,amsfonts}
\usepackage{algorithm2e}
\usepackage{graphicx}
\usepackage{textcomp}
\usepackage{xcolor}
\usepackage{makecell}
\usepackage{enumitem}
\usepackage{amsthm}
\usepackage{multirow}
\usepackage{booktabs}
\usepackage{listings}
\usepackage{graphicx}
\usepackage{array}
\usepackage{listings}
\usepackage{diagbox}
\usepackage{xcolor}  
\usepackage{subcaption}
 \usepackage{balance} 

\definecolor{customyellow}{rgb}{1.0, 1.0, 0.8}
\lstdefinelanguage{json}{
    basicstyle=\ttfamily\footnotesize,
    showstringspaces=false,
    breaklines=true,
    frame=single,
    backgroundcolor=\color{lightgray!20},
    string=[s]{"}{"},
    morestring=[b]',
    commentstyle=\color{gray},
    keywordstyle=\color{blue}
}

\graphicspath{{figs/}}
\newcommand{\tool}{\texttt{Txt2Sce}}
\def\BibTeX{{\rm B\kern-.05em{\sc i\kern-.025em b}\kern-.08em
    T\kern-.1667em\lower.7ex\hbox{E}\kern-.125emX}}
    
\begin{document}

\title{Txt2Sce: Scenario Generation for Autonomous Driving System Testing Based on Textual Reports
}


\author{
\IEEEauthorblockN{Pin Ji}
\IEEEauthorblockA{
\textit{State Key Laboratory for} \\ 
\textit{Novel Software Technology} \\
\textit{Nanjing University}\\
Nanjing, China \\
pinji@smail.nju.edu.cn}
\and
\IEEEauthorblockN{Yang Feng}
\IEEEauthorblockA{
\textit{State Key Laboratory for} \\ 
\textit{Novel Software Technology} \\
\textit{Nanjing University}\\
Nanjing, China \\
fengyang@nju.edu.cn}
\and
\IEEEauthorblockN{Zongtai Li}
\IEEEauthorblockA{
\textit{State Key Laboratory for} \\ 
\textit{Novel Software Technology} \\
\textit{Nanjing University}\\
Nanjing, China \\
lizongtai@smail.nju.edu.cn}
\and
\IEEEauthorblockN{Xiangchi Zhou}
\IEEEauthorblockA{
\textit{State Key Laboratory for} \\ 
\textit{Novel Software Technology} \\
\textit{Nanjing University}\\
Nanjing, China \\
xiangchizhou@smail.nju.edu.cn}
\and
\IEEEauthorblockN{Jia Liu}
\IEEEauthorblockA{
\textit{State Key Laboratory for} \\ 
\textit{Novel Software Technology} \\
\textit{Nanjing University}\\
Nanjing, China \\
liujia@nju.edu.cn}
\and
\IEEEauthorblockN{Jun Sun}
\IEEEauthorblockA{
\textit{School of Computing}\\
\textit{and Information Systems}\\
\textit{Singapore Management University} \\
Singapore \\
junsun@smu.edu.sg}
\and
\IEEEauthorblockN{Zhihong Zhao}
\IEEEauthorblockA{
\textit{State Key Laboratory for} \\ 
\textit{Novel Software Technology} \\
\textit{Nanjing University}\\
Nanjing, China \\
zhaozhih@nju.edu.cn}
}

\maketitle

\begin{abstract}
With the rapid advancement of deep learning and related technologies, Autonomous Driving Systems (ADSs) have made significant progress and are gradually being widely applied in safety-critical fields. 
However, numerous accident reports show that ADSs still encounter challenges in complex scenarios. 
As a result, scenario-based testing has become essential for identifying defects and ensuring reliable performance.
In particular, real-world accident reports offer valuable high-risk scenarios for more targeted ADS testing. 
Despite their potential, existing methods often rely on visual data, which demands large memory and manual annotation.
Additionally, since existing methods do not adopt standardized scenario formats (e.g., OpenSCENARIO), the generated scenarios are often tied to specific platforms and ADS implementations, limiting their scalability and portability.
To address these challenges, we propose \tool, a method for generating test scenarios in OpenSCENARIO format based on textual accident reports. 
\tool~first uses a LLM to convert textual accident reports into corresponding OpenSCENARIO scenario files. It then generates a derivation-based scenario file tree through scenario disassembly, scenario block mutation, and scenario assembly.
By utilizing the derivation relationships between nodes in the scenario tree, \tool~helps developers identify the scenario conditions that trigger unexpected behaviors of ADSs.
In the experiments, we employ \tool~to generate 33 scenario file trees, resulting in a total of 4,373 scenario files for testing the open-source ADS, Autoware.
The experimental results show that \tool~successfully converts textual reports into valid OpenSCENARIO files, enhances scenario diversity through mutation, and effectively detects unexpected behaviors of Autoware in terms of safety, smartness, and smoothness.
We further analyze the source code of Autoware in combination with its unexpected behaviors and the corresponding trigger conditions to identify specific defects in its code and design, demonstrating that \tool~can effectively help developers improve the performance of ADSs.

\end{abstract}


\section{Introduction}
Automated Driving Systems (ADSs), also known as autonomous vehicles, aim to enhance the driving experience, improve traffic safety, and alleviate road congestion~\cite{gassmann2019towards, chan2017advancements}. 
This emerging technology has garnered significant attention in both academia and industry. 
However, the highly dynamic and uncertain nature of real-world environments exposes ADSs to a wide range of threats, increasing the likelihood of system malfunctions and safety-critical failures~\cite{cui2019review,craftlaw2024}.
Studies have shown that many accidents involving ADS are attributable to latent system defects, revealing the limitations of current ADSs in reliably handling complex driving scenarios.
This underscores the pressing need for systematic testing strategies capable of evaluating ADSs under diverse environmental conditions, rare corner cases, and multi-agent interactions~\cite{tang2023survey}.

Current testing methods for ADSs primarily include real-world road testing and simulation-based testing~\cite{fremont2020formal, zhao2019assessing}.
Road testing is often time-consuming, resource-intensive, and constrained to evaluating predefined scenarios within restricted environments~\cite{haq2021can}.
Even when potentially defective scenarios are identified in real-world settings, the high variability and unpredictability of environmental factors make it difficult to ensure comprehensive and repeatable data collection.
As a result, simulation-based testing has emerged as the mainstream approach for evaluating ADS performance~\cite{kaur2021survey}.
Its primary objective is to generate critical scenarios that are likely to lead to accidents, thereby enabling rigorous and systematic validation.
By enabling the construction of complex and safety-critical scenarios that are challenging to reproduce systematically in the physical world, simulation testing significantly enhances evaluation efficiency and coverage~\cite{ding2023survey}.


However, despite the growing adoption of simulation testing, existing approaches still face critical challenges and unresolved limitations.
A major challenge in current scenario-based testing lies in the limited diversity and realism of test scenarios, largely due to the difficulty of modeling complex interactions among traffic participants. 
To address this, existing methods often rely on visual media such as images and videos to extract these interactions~\cite{zhang2023building}. 
However, constructing scenarios based on such data incurs substantial storage and annotation costs, thereby limiting scalability and automation. 
Moreover, visual data primarily capture low-level perceptual details, but lack the capacity to explicitly represent high-level behavioral semantics and causal relationships.
A further limitation arises from the lack of standardization in scenario representation.
Although OpenSCENARIO~\cite{openscenario}, a standard scenario specification, has been proposed to facilitate scenario construction, its structural and semantic complexity has hindered its widespread adoption in existing automated testing methods.
The absence of standardized representations limits cross-platform compatibility and hinders automated semantic analysis, reducing the effectiveness of testing~\cite{guo2024sovar,tang2024legend}.

To address the above challenges, we propose \tool, a method for generating OpenSCENARIO files from textual accident reports to test ADSs.
Compared to visual data, textual reports provide rich causal relationships, dynamic interactions, and contextual details without requiring extensive manual annotation, making them well-suited for scalable scenario generation.
\tool~employs carefully crafted prompts to guide a Large Language Model (LLM) in parsing accident reports and converting them into seed OpenSCENARIO files.
It then performs scenario disassembly, block mutation, and scenario reassembly to construct a scenario file tree, where each node is derived from its parent through well-defined context matching rules.
Mutation operators in \tool~include dynamic behavior mutation, trajectory adjustment, physical attribute alteration, and environmental condition variation, which together enhance the diversity and realism of the generated scenarios.
The hierarchical structure ensures that each child scenario builds upon its parent by introducing one additional scenario block.
This leads to progressively more complex and concrete scenarios.
By comparing scenarios with derivation relationships, developers can identify triggers of unexpected behaviors and better understand how specific elements affect the decisions of ADSs, aiding targeted improvements.

To validate \tool, we select the accident reports collected by the California Department of Motor Vehicles as the raw data, and all recorded accidents involve autonomous vehicles. 
We then remove duplicate reports and use \tool~to generate 33 seed scenario files, which are further expanded through mutation and assembly to produce 4,373 valid scenario files.
To measure the diversity of the generated scenarios, we use a hierarchical model to abstract and classify the generated scenarios. 
The experimental results show that \tool~generates 1,519 types of scenarios from 33 seed scenarios, effectively enhancing the richness and coverage of the generated scenarios across different environmental conditions and event combinations.
We successfully execute the generated OpenSCENARIO files in the CARLA simulator~\cite{dosovitskiy2017carla} to test Autoware~\cite{autoware_documentation}, and detect 1,788 unexpected behaviors spanning multiple categories in terms of safety,
smartness, and smoothness.
By analyzing the unexpected behaviors and their triggers, we inspect source code of Autoware to locate specific bugs, demonstrating the effectiveness of \tool~in aiding ADS improvement.
The main contributions of this paper are as follows:
\begin{enumerate}
    \item \textbf{Method.} We propose an OpenSCENARIO file generation method \tool~that can convert textual accident reports into OpenSCENARIO format. \tool~can further generate a scenario file tree through mutation and assembly based on the converted seed scenario files. The derivation relationships within the scenario tree effectively assist developers in locating defects and accelerating the performance improvement of ADSs.
    \item \textbf{Tool. }We implement the \tool~using Python and a state-of-the-art large language model, and release it as an open-source tool. This tool enables developers to rapidly generate a large number of diverse, standardized scenario files from natural language descriptions.
    \item \textbf{Study.} We use \tool~to convert 33 distinct OpenSCENARIO files and generate 4,373 OpenSCENARIO files to test Autoware. These scenarios reveal various unexpected behaviors of Autoware related to safety, smoothness, and smartness across five categories. By analyzing these behaviors, we identify specific code and design defects within Autoware, demonstrating the effectiveness of \tool~in improving the performance of ADSs.
\end{enumerate}

\section{Background}
\subsection{Autonomous Driving System}
\begin{figure}[h]
  \centering
  \includegraphics[width=\linewidth]{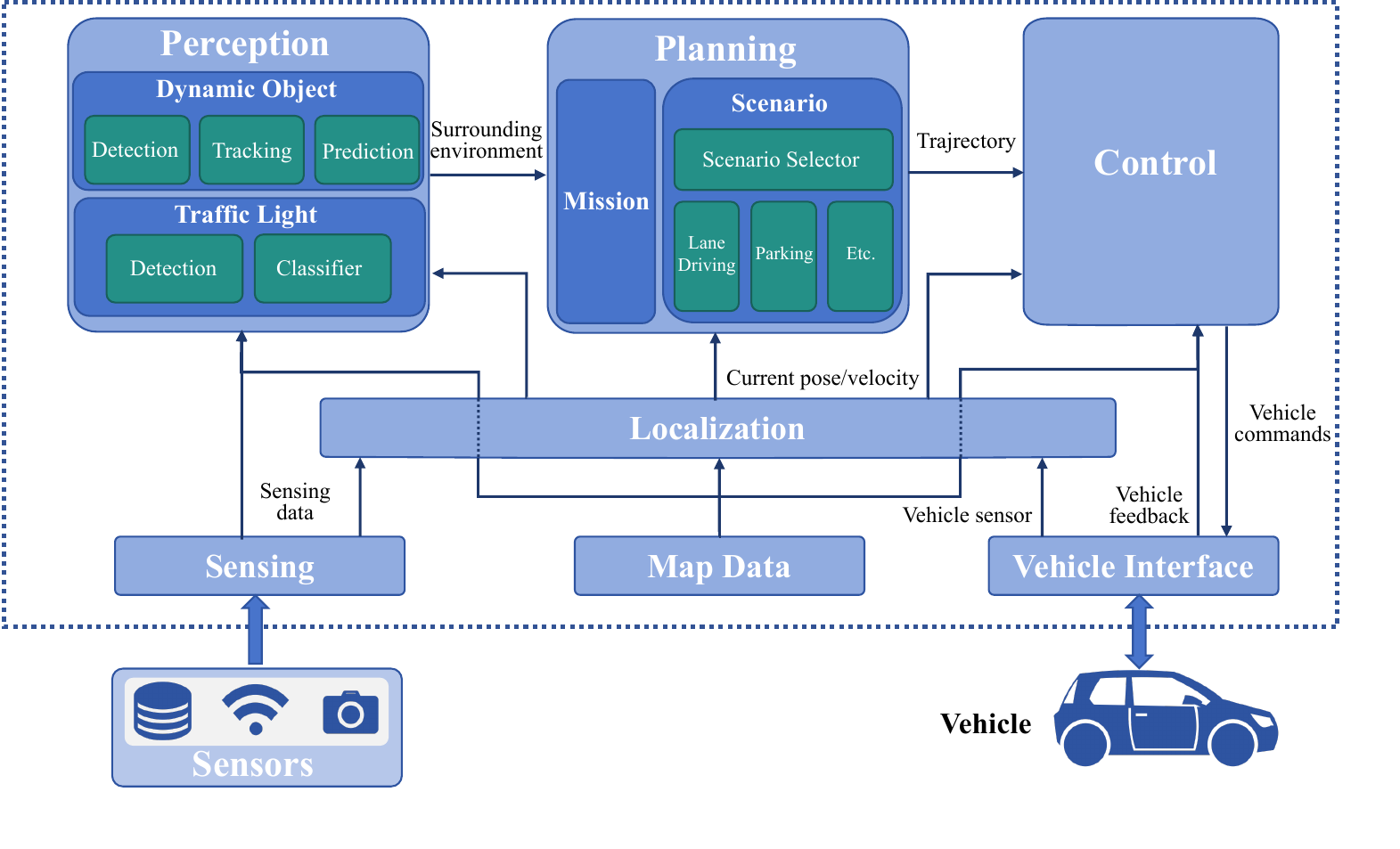}
  \caption{Architecture Diagram of Autoware}
  \label{fig:autoware}
  \vspace{-1.5em}
\end{figure}

Autonomous Driving Systems (ADSs) are complex intelligent systems that integrate perception, planning, control, and localization to enable vehicles to respond in real time to dynamic environments and navigate safely without human intervention~\cite{yurtsever2020survey, pendleton2017perception}.
With the rapid development of autonomous driving, various system architectures have been proposed.
To illustrate a representative structure, we refer to Autoware—an open-source ADS stack—which organizes its core modules into perception, localization, planning, and control, as shown in Figure~\ref{fig:autoware}~\cite{kato2018autoware}.
In Autoware, the perception module collects environmental data from various sensors, identifies surrounding dynamic objects, and tracks their movements to predict future trajectories, thereby enabling safe navigation in complex traffic environments~\cite{autoware_documentation}.
The localization module fuses map data with sensor inputs to estimate the vehicle’s current position and velocity, providing critical support for downstream path planning.
The planning module generates optimal driving trajectories by integrating perception and localization information, enabling the vehicle to select appropriate routes and driving behaviors. 
It also supports flexible mode switching across diverse driving scenarios—such as lane keeping, traffic signal handling, and automated parking—to ensure appropriate decision-making under varying road conditions.
Finally, the control module converts planned trajectories into low-level control commands—such as steering, acceleration, and braking—and uses real-time feedback to ensure the vehicle accurately follows the intended path.
Through the seamless interaction of these modules, Autoware realizes an end-to-end autonomous driving workflow.

\subsection{Autonomous Driving Test Scenario}
Autonomous driving test scenarios simulate real-world conditions—such as traffic, weather, and interactions among road users—to evaluate ADS performance under controlled and repeatable settings~\cite{stadler2022credibility, ren2022survey}. 
To standardize scenario creation, the Association for Standardization of Automation and Measurement Systems (ASAM) introduces OpenSCENARIO, a specification for describing dynamic scenarios in ADS testing~\cite{openscenario}. 
It captures traffic participant behaviors and supports static elements like weather, signals, and infrastructure. OpenSCENARIO is highly extensible and integrates with standards like OpenDRIVE~\cite{opendrive} and OpenCRG~\cite{opencrg} for road modeling. 
Mainstream simulators such as CARLA~\cite{dosovitskiy2017carla}, PreScan~\cite{wang2021simulation}, and CarMaker~\cite{ipg_carmaker} offer native support. 
The OpenSCENARIO specification comprises several components: parameter declarations define reusable scenario-wide variables; catalog directories specify paths to predefined elements such as vehicles, obstacles, and actions; the road network module references external road models like OpenDRIVE; entities describe the physical and behavioral characteristics of both dynamic and static participants; and the storyboard outlines the temporal flow of the scenario, including actions, conditions, and transitions to ensure logical and realistic testing.


\subsection{The motivation of \tool}
Despite growing interest in scenario-based testing, existing methods still face key limitations in practice.
Specifically, these limitations include: 

\textbf{Limitation 1: Lack of Realism and Diversity in Dynamic Scenario Construction.}
Modeling realistic interactions among traffic participants—such as vehicles, pedestrians, and cyclists—remains a major challenge in dynamic scenario construction~\cite{li2021scegene,ding2023survey}.
To approximate such dynamic interactions, many existing methods rely on visual media—such as traffic accident images and videos—to extract behavioral cues for constructing  scenarios~\cite{tang2023survey}.
Representative datasets used in these approaches include KITTI~\cite{geiger2013vision}, Udacity~\cite{srinivasanramanagopal2018failing}, and GTSRB~\cite{stallkamp2012man}.
However, vision-based approaches suffer from two major drawbacks.
First, while visual data contain low-level perceptual features, they lack the capacity to explicitly represent high-level semantic behaviors and causal relationships among traffic participants.
Second, extracting semantics from visual data is computationally expensive and annotation-heavy, limiting the efficiency and scalability of large-scale scenario generation.
Additionally, the behavioral variety expressed in generated scenarios is constrained by the coverage and granularity of the source datasets.
\textbf{Limitation 2: Lack of Standardization in Scenario Representation.}
Previous studies have summarized representative scenario description languages and analyzed their adoption in ADS testing~\cite{ma2021traffic, tang2023survey, riedmaier2020survey}.
These studies show that most scenario description languages—including established ones like OpenSCENARIO and Scenic—are not widely adopted in existing testing methods, primarily due to the semantic complexity involved in describing dynamic and context-rich scenarios~\cite{tang2023survey}. 
The lack of a unified scenario representation standard in existing methods has resulted in fragmented scenario formats, with each method defining its own structure and semantics. This inconsistency makes it difficult to extend, reuse, or integrate scenarios across simulation platforms and tool-chains. Without a common standard like OpenSCENARIO, the scenarios are often tied to specific simulators or testing tools, reducing their interoperability. Furthermore, the absence of standardized structure limits the application of automated semantic analysis techniques, thereby weakening the interpretability and generalizability of testing results.

To address the aforementioned limitations, we propose \tool, a test scenario generation method for ADS testing that leverages textual accident reports.
Compared to visual data, textual accident reports efficiently convey causal relationships, dynamic interactions, and contextual details through natural language—without requiring extensive manual annotation—making them a more scalable and effective source for generating large volumes of realistic, behaviorally rich test scenarios.
The scenarios generated by \tool~are described in OpenSCENARIO, an international standard for ADS testing supported by major automotive manufacturers. Based on its syntax specification, we design methods for scenario conversion, disassembly, block mutation, and assembly to generate diverse scenarios with derivation relationships.

\section{Approach}
\begin{figure*}[h]
  \centering
  \includegraphics[width=0.9\linewidth]{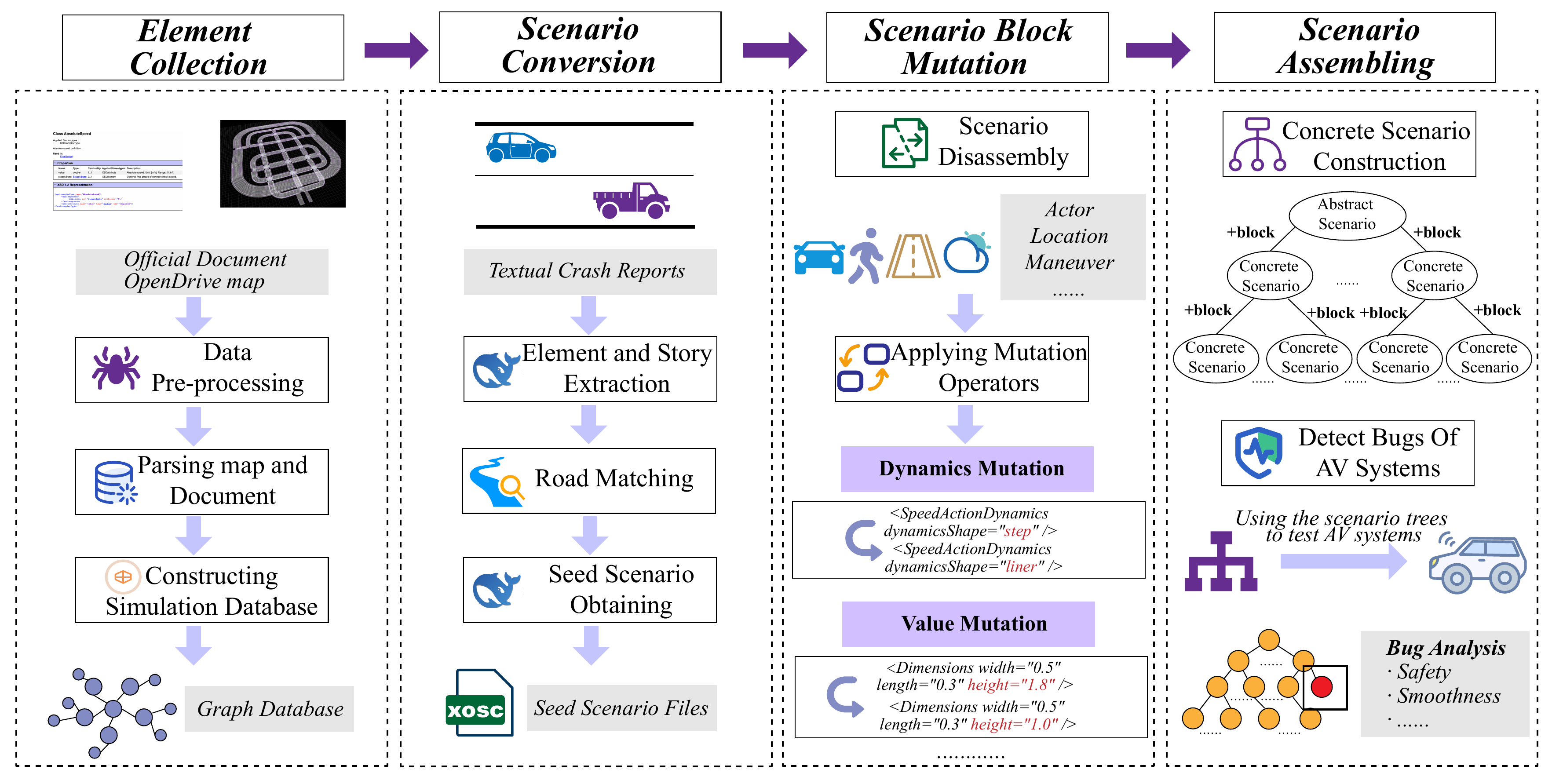}
  \caption{The Overview of \tool}
  \label{fig:overview}
  \vspace{-2.0em}
\end{figure*}

In this section, we present the design and implementation details of \tool, which aims to convert textual accident reports into corresponding seed scenarios and further expand them into diverse scenarios for testing ADSs. As shown in Figure~\ref{fig:overview}, it mainly includes the following steps: 

\begin{enumerate}
    \item \textbf{Scenario Element Collection:} \tool~constructs an entity database from the OpenSCENARIO specification and a map database from the OpenDRIVE file, both stored in graph structures to support efficient scenario generation and mutation.

    \item \textbf{Scenario Conversion:} \tool~employs a large language model (LLM) to extract key scenario elements from textual reports and convert them into OpenSCENARIO-compliant seed files, ensuring semantic and spatial consistency with real-world scenarios.

    \item \textbf{Scenario Block Mutation:} To enhance the diversity of generated scenarios, \tool~disassembles seed scenarios into semantic blocks and applies mutation operators across four categories: dynamics, trajectories, physical attributes and environment.

    \item \textbf{Scenario Assembly:} Mutated blocks are reassembled into valid scenario files following a hierarchical structure. This enables the generation of derivation trees and supports behavior-trigger analysis through comparisons between scenarios with derivation relationships.

\end{enumerate}


\subsection{Scenario Element Collection}

This step focuses on collecting essential elements required for scenario construction, primarily including scenario specification data and map structure information, which are organized into graph databases to support subsequent generation tasks.

\subsubsection{Construct Entity Database from OpenSCENARIO Documentation}  
We use \textit{Scrapy} to extract UML-based OpenSCENARIO documentation, where each scenario element is defined as a class, enumeration, or primitive type, with metadata such as name, type, cardinality, and description.
Based on this specification, we build a graph-based entity database with two node types: \textit{XSDElement} and \textit{XSDAttribute}.  
\textit{XSDElement} represents major scenario components such as weather, vehicles, and actions, which form the core structure of OpenSCENARIO-based XML descriptions.  
\textit{XSDAttribute} nodes capture fine-grained element properties that are atomic and not further decomposable.  
Three types of edges are used: containment between elements, inheritance between types, and has-attribute relationships.  
This database encodes the relationships among scenario elements, attribute types, and valid value ranges, serving as a schema-level reference for scenario generation.

\subsubsection{Construct Map Database from OpenDRIVE Files} 
To support spatial reasoning in scenario generation, we parse OpenDRIVE files and construct a graph-based map database that encodes road topology and lane-level semantics.
Each node in the graph corresponds to a road segment, with attributes including \textit{name}, \textit{length}, and lane metadata (e.g., \textit{lane\_ids}, \textit{lane\_types}, \textit{lane\_directions}, \textit{lane\_change} permissions).
Directed edges capture topological connections between segments, annotated with junction-level information such as \textit{connecting\_road}, \textit{junction\_id}, and lane transitions (e.g., \textit{start\_lane\_id}, \textit{end\_lane\_id}).
Additional edge attributes include connection geometry  and infrastructure context (e.g., \textit{traffic\_light\_x}, \textit{traffic\_light\_y}).
This map database provides the spatial context for locating entities and supports downstream tasks such as road matching, scenario positioning, and trigger condition validation.
Together with the scenario entity database, it forms the semantic-spatial foundation of our scenario generation pipeline.


\subsection{Text Report Conversion}

\begin{figure*}[h]
    \centering
    \begin{subfigure}{0.24\textwidth}
        \centering
        \includegraphics[width=\linewidth]{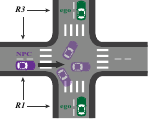}
        \caption{Relationship $R1$ and $R3$}
        \label{fig:sub1}
    \end{subfigure}
    \hfill
    \begin{subfigure}{0.24\textwidth}
        \centering
        \includegraphics[width=\linewidth]{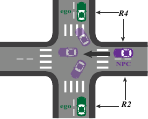}
        \caption{Relationship $R2$ and $R4$}
        \label{fig:sub2}
    \end{subfigure}
    \hfill
    \begin{subfigure}{0.24\textwidth}
        \centering
        \includegraphics[width=\linewidth]{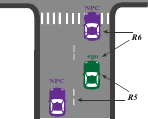}
        \caption{ Relationship $R5$ and $R6$}
        \label{fig:sub3}
    \end{subfigure}
    \hfill
    \begin{subfigure}{0.24\textwidth}
        \centering
        \includegraphics[width=\linewidth]{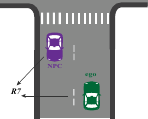}
        \caption{Relationship $R7$}
        \label{fig:sub4}
    \end{subfigure}
    
    \caption{Example Figures For Seven Relative Position Relationships}
    \label{fig:position-main}
    \vspace{-2.0em}
\end{figure*}

\subsubsection{Prompt Design For Parsing Reports}
A Large Language Model (LLM) is capable of contextual understanding and generating natural language based on extensive training data~\cite{lai2024survey, chang2024survey}.
We leverage a LLM (i.e., DeepSeek) to extract key scenario elements from the accident reports collected by the California Department of Motor Vehicles (DMV), which describe AV-involved incidents in natural language. 
To fully utilize the LLM’s capacity for long-context comprehension, we design a multi-turn, dialogue-style prompt that progressively guides the extraction of scenario information.
This interactive prompting strategy allows iterative refinement, improving accuracy and completeness~\cite{zhang2020dialogpt, achiam2023gpt}.
We define interacting traffic participants as Non-Playable Characters (NPCs), and static objects as obstacles.
For each, we extract their quantity, type, location, and behavior from the report text. 
In order to  ensure the realism of generated scenarios, we design prompts that transform pre-crash NPC behaviors into ordered event sequences composed of 13 fine-grained actions(e.g., turns, acceleration, deceleration).
This standardization improves output consistency and ensures logical continuity in generated scenarios.
As shown in  Figure~\ref{fig:position-main}, to determine the positions of NPCs, we prompt the LLM to convert absolute road positions into seven types of relative positions with respect to the Autonomous Vehicle (AV).
For cases in $R5$–$R7$, the prompt also determines lane-level alignment.
This strategy decouples behaviors from road layouts, supporting reuse across various junction types.
Prompts  and implementation details are provided in our public repository.

\subsubsection{Position Alignment and Road Segment Matching}

To determine the spatial configuration of the AV and NPCs, \tool~performs a two-step process: road segment filtering and initial lane assignment.
In the first step, \tool~extracts high-level scenario constraints from the accident report, such as the number and type of lanes, directional connectivity (e.g., left turns), and relative positions of NPCs (e.g., R1–R7). It then queries the pre-constructed map database to eliminate road segments that do not satisfy these constraints. One suitable road or junction is randomly selected from the remaining candidates to serve as the spatial context.
In the second step, \tool~assigns initial lanes to both the AV and NPCs. By default, vehicles are placed in the leftmost available lane, unless overridden by behavior-specific requirements. The assignment strategy varies depending on whether the AV and NPCs are located on the same road segment:




\emph{Case 1: NPC is not on the same road segment as the AV (R1–R4, R7).}  
The lane is chosen based solely on the NPC’s behavior sequence. \tool~scans the sequence to identify the first action requiring a lane constraint (e.g., a left turn or a lane change).  
For example, a sequence like “ad” requires the NPC to start in the leftmost left-turn lane, whereas a sequence like “afd” necessitates starting in a more rightward lane to support a later left-lane-change maneuver.

\emph{Case 2: NPC shares the same road segment as the AV (R5, R6).}  
When the NPC and AV are on the same road segment, \tool~first determines the AV’s lane, as it influences the behaviors of NPC feasibility. Although the AV defaults to the leftmost lane, it is reassigned to the rightmost feasible lane under the following conditions: (1) the NPC overtakes or passes from the left; (2) the NPC is in the same lane and changes to the left; (3) the NPC is in a different lane and changes to the right. Once the lane of AV is assigned, \tool~determines the NPC’s lane based on whether it shares the AV’s lane or operates independently.


\subsubsection{LLM-based OpenSCENARIO Generation}

After completing position alignment and road matching, \tool~generates OpenSCENARIO-compliant seed files using a hybrid approach that combines template-based structure filling with LLM-driven content generation.
This process comprises two stages:
First, \tool~initializes a general scenario template based on the selected map. This includes parsing the template structure, configuring the road network, annotating traffic signals, and computing entity positions—forming a static scenario skeleton with semantic slots reserved for subsequent generation.
In the second stage, \tool~employs the LLM to fill in all semantic-level components, such as object definitions (e.g., type, color), initial speeds, signal operations, speed changes, trajectories, and event sequences of NPCs. The LLM takes structured inputs—entity positions, behavior constraints, and interaction relationships—and generates valid OpenSCENARIO XML fragments, which are then inserted into the template to complete the scenario file.
Compared with traditional rule-based or template-based approaches, this LLM-integrated method offers three advantages:
(1) It eliminates the need for hard-coded logic to handle behavioral combinations, improving scalability and generality;
(2) It accepts diverse input formats without requiring additional preprocessing;
(3) It ensures contextual coherence in generated behaviors, producing semantically rich and realistic scenarios.

\subsection{Scenario Block Mutation}
To enhance the diversity of generated scenarios, we design a scenario disassembly method $\delta$ and a set of scenario block mutation operators.
The disassembly method identifies independently mutable scenario blocks from a seed scenario file, producing two outputs:
(1) a scenario template $\mathbf{T}$ that retains the static structure and defines block insertion positions, and
(2) a set of mutable blocks $\mathbb{B} = {b_1, b_2, \dots, b_n}$.
The block types cover both environmental elements—such as weather, traffic signals, and obstacles—and core dynamic elements—such as NPC definitions and their associated event sequences.
Each individual event within an event sequence is further treated as a separate block to support fine-grained mutation.
In practice, blocks are extracted based on their XML tag names and name attributes, leveraging the predefined syntax structure of OpenSCENARIO.

To account for the diverse semantics and characteristics of different types of scenario blocks, we design multiple families of mutation operators, denoted as $\mathbf{M}$, which include:
\begin{enumerate}[left=0pt]
\item \textbf{Dynamics Mutation}: This family of mutation operators targets the dynamic characteristics of moving entities to simulate diverse driving behaviors and motion patterns.
The specific operators are as follows: 
\begin{itemize}
    \item \textit{Target Speed Mutation (TSM)}: Alters the \textit{AbsoluteTargetSpeed} to vary how fast entities move.
    \item \textit{Dynamic Transition Mutation (DTM)}: Mutates \textit{TransitionDynamics} attributes such as \textit{value} and \textit{dynamicsShape}.
    \item \textit{Vehicle Performance Mutation (VPM)}: Modifies performance limits like \textit{maxAcceleration}, \textit{maxDeceleration}, and \textit{maxSpeed}.
\end{itemize}

\item \textbf{Trajectory Mutation}: This category of operators adjusts the motion trajectories of NPCs and the initial position of the ADS to introduce spatial diversity while preserving the original event semantics. 
\tool~adopts the Waypoint Mutation (WPM) operator, which perturbs the lateral offset values in lane-change-related \textit{Waypoint} elements to simulate variations in driving paths.

\item \textbf{Physical Attribute Mutation}: This type of mutation operator refers to mutations that involve altering the physical properties of an entity. It includes: 
\begin{itemize}
    \item \textit{Dimension Mutation (DM)}: Adjusts width, length, and height to simulate size variations of vehicles, pedestrians, or obstacles.
    \item \textit{NPC Category Mutation (NCM)}: Changes the category of an NPC (e.g., from car to motorbike), thereby altering its shape, dynamics, and the simulation model used.


\end{itemize}

\item \textbf{Environment Mutation}: This family of mutation operators modifies external environmental conditions that may affect the behavior of the ADS. The specific operators include:
\begin{itemize}
    \item \textit{Weather Mutation (WM)}: Alters weather type and generates related parameters (e.g., visibility, sun intensity, azimuth, elevation, precipitation). The road friction coefficient is adjusted accordingly.
    
    \item \textit{Traffic Signal Mutation (TSM)}: Changes traffic light states (i.e., \textit{TrafficSignalState}) during interactions to test ADS behavior under different signal conditions.

    \item \textit{Obstacle Insertion Mutation (OIM)}: Inserts static obstacles relative to the AV's path, with randomized physical attributes (length, width, height) to simulate potential hazards.



\end{itemize}
\end{enumerate}

To determine the final values of mutated attributes, \tool~adopts three value selection strategies based on the attribute type and scenario semantics:
(1) Random Sampling: Uniform sampling within predefined ranges, used for discrete or bounded attributes;
(2) Gaussian Mutation: Adds controlled noise drawn from a normal distribution, suitable for continuous values;
(3) Context-Aware Computation: Computes values dynamically based on map topology and scenario context.


\subsection{New Scenario Generation and Application}
\subsubsection{Scenario Assembly}
We design a scenario assembly method $\alpha$ that inserts mutated scenario blocks into a scenario template to generate semantically richer scenarios. In \tool, blocks are inserted in a fixed order: weather, NPCs, traffic signals, events, and obstacles. As shown in Figure~\ref{fig:overview}, this results in a hierarchical scenario tree, where each node extends its parent with one additional block.
The total number of generated scenarios is determined by the number of mutations for each block:
\begin{align}
    Sum_{sce} &= \mathbf{M}(b_1) + \mathbf{M}(b_1) \times \mathbf{M}(b_2) + \dots + \prod_{i=1}^{n} \mathbf{M}(b_i)
\end{align}
where $\mathbf{M}(b_i)$ is the number of mutated versions of block $b_i$.

This hierarchical design allows developers to trace the behavioral impact of each added block by comparing parent-child pairs or sibling pairs. To improve testing efficiency, \tool~applies a diversity-aware pruning strategy: when multiple child scenarios are semantically redundant, only representative ones are retained.
The pruning strategy is guided by heuristics based on block type, parameter distance, and structural similarity—defined by the number and position of entities and the composition of event sequences. This helps reduce redundancy while preserving semantic coverage.

\subsubsection{Bug Detection}

In this paper, we focus on the unexpected behaviors of the ADS in three dimensions: safety, smoothness, and smartness. 

\textbf{Safety.}  
To evaluate the ADS's ability to avoid hazards, we detect collisions based on sudden deceleration, a widely adopted indicator in prior studies~\cite{sequeira2020evaluation}.
Runtime metrics such as speed and braking are automatically recorded, and a collision is flagged when the measured \textit{jerk} $j$ exceeds a predefined threshold $T_j$. 
\textit{Jerk} reflects rapid changes in acceleration and is defined as:

\begin{equation}
a = \frac{dv}{dt}, \quad j = \frac{da}{dt}
\end{equation}

where $dv$ is the change in velocity (m/s), $da$ is the change in acceleration (m/s$^2$), and $dt$ is the time interval (s).
Following the prior study~\cite{sequeira2020evaluation}, jerk is computed over short intervals (5–15 ms) to capture sudden deceleration events.

\textbf{Smoothness.}  
Smoothness reflects the ADS’s ability to operate naturally and comfortably, ensuring both passenger comfort and vehicle stability~\cite{xiang2022comfort}.
We use two indicators: \textit{jerk} to measure longitudinal stability and \textit{yaw rate} to capture lateral motion smoothness.
To account for short-term fluctuations, both are smoothed using a 1-second moving average:

\begin{equation}
\operatorname{jerk}_{\text{avg}}(t) = \frac{1}{N} \sum_{i=t-N+1}^{t} \operatorname{jerk}_i
\end{equation}

\begin{equation}
\dot{\psi}(t) = \frac{d\psi(t)}{dt}, \quad
\text{Yaw}_{\text{avg}}(t) = \frac{1}{N} \sum_{i=t-N+1}^{t} \dot{\psi}_i
\end{equation}

\textbf{Smartness.}  
Smartness captures the reasoning ability of the ADS when facing complex situations. We define it by identifying high-level failures that indicate deficiencies in scenario understanding or decision-making: (1) failure to start: inability to begin motion after the initial state; (2) failure to reach the goal: failing to complete the scenario route; (3) failure to interpret signals/obstacles: ignoring traffic lights or colliding with visible static objects.

\section{EXPERIMENT DESIGN}
We implement \tool~in Python, using DeepSeek (i.e., \textit{deepseek-chat}) for natural language understanding and Neo4j to store the graph-based scenario databases.  
All experiments are conducted on a desktop running Ubuntu 20.04, equipped with an Intel Core i7-14700KF CPU and an NVIDIA RTX 3070 Ti GPU.  
For simulation, we use CARLA~\cite{dosovitskiy2017carla} with \textit{scenario\_runner}~\cite{scenario_runner} as the OpenSCENARIO-compatible execution engine.  
The scenario trees generated by \tool~are subsequently used to test Autoware.ai.

We do not include baseline methods because, to the best of our knowledge, no existing approach can directly generate standardized OpenSCENARIO files from textual accident reports.  
Methods such as SoVAR~\cite{guo2024sovar} and LeGEND~\cite{tang2024legend} are tightly coupled with specific simulators (e.g., LGSVL) and produce perception-layer data instead of reusable, platform-independent scenario files.  
This makes them incompatible with our OpenSCENARIO-based testing workflow and evaluation criteria.
We evaluate the performance of \tool~through the following research questions:

\begin{itemize}
    \item \textbf{RQ1. Quality:} Do the textual descriptions and the seed scenario files generated by \tool~convey equivalent semantics? Can \tool~enhance the diversity of generated test scenarios?

    \item \textbf{RQ2. Effectiveness:} How many unexpected behaviors can \tool~reveal in ADS testing? How diverse are these behaviors?

    \item \textbf{RQ3. Bug Study:} Can the identified unexpected behaviors help developers locate concrete defects in the ADS?
\end{itemize}

\subsection{The Source of Textual Reports}
The accident reports used in this study are collected from the California DMV, where all documented incidents involve autonomous vehicles.  
Since 2014, the DMV has mandated permit-holding AV testing organizations to submit standardized accident reports.  
These reports consist of five sections, with the final part offering a natural language description detailing pre-crash behaviors~\cite{song2021automated}.  
We collect 387 reports published between 2019 and April 2024 from the DMV's official website~\cite{DMV}, all in PDF format.  
To extract relevant textual content, we use PDF parsing tools (e.g., PDFMiner~\cite{pdf_parser}) to isolate the free-text portion describing vehicle interactions.  
Compared to NHTSA reports~\cite{nhtsa2024crashsurvey}, which often blend structured and narrative data and mostly focus on traditional vehicle crashes, DMV reports separate behavioral narratives clearly and concentrate on autonomous vehicles, enhancing both data clarity and domain relevance for ADS testing.
After extraction, we filter out rear-end collisions by keyword search and pass the remaining 180 reports into the LLM for interpretation.  
\tool~then generalizes these reports based on core scenario components—such as NPC count, relative positions, event sequences, and obstacle configurations—enabling compositional mutation from representative seeds.
Rather than treating each report as a distinct case, we cluster structurally similar descriptions and retain one representative scenario per group to improve testing efficiency while maintaining scenario diversity.  
We further exclude low-complexity cases lacking both NPCs and obstacles, as well as seeds whose event sequences are subsets of others.  
Following this process, 33 representative seed scenarios are selected for compositional generation.

\subsection{Parameters Settings}
In our experiments, we configure three sets of parameters: those for seed scenario generation, scenario block mutation, and test oracles for detecting unexpected ADS behaviors.
During seed generation, default values are assigned to parameters not explicitly specified in the accident reports to ensure scenario completeness. 
As the scenario assembly process constructs a tree where each mutation introduces a new branch, the number of generated scenarios grows exponentially. 
To balance diversity and computational cost, we apply two mutations per block, except for weather blocks, traffic light states, and NPC categories, where all possible values are enumerated. 
This yields an approximate binary expansion at most layers.
Table~\ref{tab:mutation_parameters} summarizes the default values for seed generation, mutation ranges, and value selection strategies, where $x$ denotes the original (pre-mutation) value.
Due to space limitations, the complete parameter settings table is available in our repository.
For the collision detection threshold $T_{cj}$, we use $\pm 300~\mathrm{m/s^3}$ based on prior work.
Following prior studies~\cite{martin2008investigation, nguyen2019insight}, we set the jerk interval $I_{sj}$ and yaw rate interval $I_{yr}$ to $[0, 0.9]~\mathrm{m/s^3}$ and $[-10^\circ/\mathrm{s}, 10^\circ/\mathrm{s}]$, respectively, for evaluating smoothness.

\begin{table*}[h]
\centering
\footnotesize
\caption{Parameter Settings for Seed Generation and Mutation Operators}
\label{tab:mutation_parameters}
\begin{tabular}{c|c|p{2.5cm}|p{2.5cm}|p{4cm}|p{3cm}}
\toprule
\textbf{Type} & \textbf{Operator} & \textbf{Parameter} & \textbf{Default Value}&\textbf{Value Range} &\textbf{Mutation Pattern} \\
\hline
 \multirow{7}{*}{\makecell{\textbf{Dynamics}\\\textbf{Mutation}}} & TSM & target speed ($\mathrm{m} / \mathrm{s}$)& sedan: 6 bicycle: 3 ... &acc: [$x$, 1.5$x$] dec: [0.5$x$, $x$] & randomly sampled\\
 \cline{2-6}
 &  \multirow{3}{*}{DTM} & dynamics shape & `linear' &`cubic', `sinusoidal', `linear' & randomly sampled
\\
 \cline{3-6}
 & & value & 1 &[1,10] & randomly sampled \\
\cline{2-6}
 & \multirow{3}{*}{\makecell{VPM}} &max acceleration& 10 m/s& - & Gaussian Mutation\\
 \cline{3-6}
 & & max deceleration & 10 m/s &- & Gaussian Mutation\\
 \cline{3-6}
 & & max speed & 70 m/s & - & Gaussian Mutation\\
 \cline{1-6}
 \multirow{2}{*}{\makecell{\textbf{Trajectory}\\\textbf{Mutation}}} &\multirow{2}{*}{WPM}& offset & 0 &[-$length_{road}$/2, $length_{road}$/2] & randomly sampled\\
 \cline{3-6}
 & & Initialization location & report-dependent & map-dependent & Context-aware Mutation\\
\cline{1-6}
 \multirow{4}{*}{\makecell{\textbf{Physical}\\\textbf{Attribute}\\\textbf{Mutation}}} & \multirow{3}{*}{DM} & width(m) &sedan: 1.8 van: 2.1 ... & - & Gaussian Mutation\\
 \cline{3-6}
 & & length(m) & sedan: 4.5 van: 5.3 ...& - & Gaussian Mutation\\
 \cline{3-6}
 & & height(m) & sedan: 1.5 van: 1.8 ...& - & Gaussian Mutation\\
 \cline{2-6}
 & NCM & category & report-dependent & `sedan', `bicycle', `van', ...& Enumerative Mutation
\\
 \cline{1-6}
 \multirow{8}{*}{\makecell{\textbf{Environment}\\\textbf{Mutation}}} & \multirow{7}{*}{WM}& precipitation intensity & randomly sampled &[0.5, 1] & randomly sampled\\ 
 \cline{3-6}
 & &visibility& randomly sampled &rainy: [100,500] fogy:[50,100]... & randomly sampled \\
 \cline{3-6}
 & &azimuth& randomly sampled &$[0,2 \pi]$ & randomly sampled \\
 \cline{3-6}
 & &solar elevation angle& randomly sampled &$[-\pi, \pi]$ & randomly sampled \\
 \cline{3-6}
 & &friction coefficient& randomly sampled &rain:[0.2,0.5] fog:[0.5,0.8]... & randomly sampled \\
 \cline{2-6}
 & TSM & state &`off' &`green',`yellow',`red'& Enumerative Mutation \\
 \cline{2-6}
  & OIM & position & report-dependent & map-dependent & Context-aware Mutation \\
\bottomrule
\end{tabular}
\vspace{-2.0em}
\end{table*}

\subsection{Pruning Strategy Configuration}
\label{sec:pru-setting}

To support diversity-aware pruning during scenario tree generation, we design a hierarchical abstraction model that mirrors the scenario assembly structure, including weather, traffic signals, NPCs, event sequences, and obstacles. This model encodes key structural attributes to identify similar scenarios for pruning. To improve efficiency and reduce dimensional redundancy, not all mutated attributes are treated equally. Numerical attributes (e.g., azimuth, elevation, visual range, target speed) are discretized using uniform binning with five bins per attribute. Enumerated attributes (e.g., signal states, NPC categories) are grouped by symbolic values. Semantic behavior features—such as acceleration styles and event sequences—are categorized based on high-level driving semantics. Each scenario is then encoded into a structural feature vector, enabling fast clustering. From each cluster, we randomly retain 50\% of the scenarios to reduce redundancy while preserving semantic diversity.

\section{Result and Evaluation}
\subsection{RQ1: The Quality of tests generated by \tool}

To evaluate the quality of the generated scenarios, we focus on two aspects of the scenarios: authenticity and diversity. 
Authenticity is evaluated in two steps: first, checking whether the LLM's outputs accurately reflect the descriptions in the accident reports; second, verifying whether the generated scenario files semantically align with the original reports, including key aspects such as participants, positions, and behavioral events.
We conduct the authenticity evaluation through a manual review process performed by the three authors of this study.
Each author independently evaluates both the correctness of the responses of the LLM to each question and the accuracy of the simulation of the generated seed scenario files.
In cases of disagreement, we employ discussion and consensus-based resolution to ensure objectivity. 
For diversity, we examine whether the scenario tree produced through mutation and assembly effectively increases variation across environment settings and interaction patterns, enabling broader behavioral coverage for ADS testing.

\textbf{Authenticity.} 
Table~\ref{tab:llm_eva} presents the accuracy of DeepSeek in answering key scenario construction questions from accident reports: NPC type and count, their relative positions, behavior sequences, and obstacle information.
DeepSeek performs well on questions involving concrete or countable data (e.g., NPC/obstacle types and quantities), with accuracy exceeding 95\%.
Accuracy slightly decreases for NPC positions (87.8\%) and behavior sequences (90.2\%), due to the implicit, context-dependent nature of these descriptions. 
The reduced accuracy stems from inconsistent writing styles and occasional omissions of key spatial or behavioral details.
Obstacle-related questions yield higher accuracy, as static objects are easier to localize and interpret than dynamic NPCs with evolving actions.
We manually correct the few extraction errors and use \tool~to generate seed scenario files, validating them through execution in CARLA.
All 33 seeds run successfully, and 93.9\% align semantically with the original accident descriptions.
The two mismatches are due to timing misalignments in NPC behaviors, which can be addressed by further mutation (e.g., adjusting speed or timing) to enrich behavioral variations and recover intended interactions.

\begin{table}[h]
\centering
\footnotesize
\caption{The Accuracy of the LLM's Outputs}
\label{tab:llm_eva}
\begin{tabular}{lccccc}
\toprule
\textbf{Category} & \textbf{Quantity} & \textbf{Type} & \textbf{Position} & \textbf{Events} \\
\midrule
NPC & 96.9\% &97.6\%&87.8\%&90.2\%\\
Obstacle & 96.9\% & 96.9\%&96.9\%& N/A\\
\bottomrule
\end{tabular}
\vspace{-1.0em}
\end{table}

\textbf{Diversity.} 
We adopt the similar semantic categorization scheme used in the pruning phase, classifying the generated scenarios based on all mutation dimensions listed in Table~\ref{tab:mutation_parameters}, in order to quantify the diversity achieved by~\tool.
Specifically, \tool~generates a total of 4,373 new and valid scenario files based on 33 seed scenario files. 
To quantitatively demonstrate the ability of \tool~to generate diverse scenario variants, we analyze mutation results across multiple abstraction levels. We focus on three key abstraction levels: weather, event sequence, and NPC configuration. 
The number of abstract representations is obtained from the seed scenarios, with only one abstract weather due to the default sunny setting.
As shown in Table~\ref{tab:diversity_variant_count}, even with a limited number of abstract representations, \tool~can derive a large number of diverse concrete variants at each level through compositional mutation.
Considering all combined mutation dimensions, a total of 1,519 unique scenario categories are formed.
The experimental results demonstrate that the scenario block mutation and scenario assembly methods in \tool~effectively enhance the diversity of scenarios, encompassing a wide range of scenario combinations and providing comprehensive coverage for evaluating ADS.


\begin{table}
\centering
\caption{Number of Scenario Variants Generated}
\label{tab:diversity_variant_count}
\begin{tabular}{lcc}
\toprule
\makecell{\textbf{Abstraction}\\\textbf{Level}} & \makecell{\textbf{Number of Abstract}\\\textbf{Representations}} & \makecell{\textbf{Number of}\\\textbf{Variants Generated}} \\
\midrule
Weather& 1 & 76 \\
Event Sequence & 33 & 662 \\
Entity Configuration & 7 & 264 \\
\bottomrule
\end{tabular}
\vspace{-2.0em}
\end{table}

\subsection{RQ2: The Effectiveness of tests generated by \tool}
After applying the pruning strategy described in Section~\ref{sec:pru-setting}, a total of 1,832 runnable scenario files are retained. In RQ2, we use these scenarios to test Autoware.
We classify unexpected ADS behaviors into five categories:
(1) \textbf{Failure to Start}: the vehicle fails to move despite having a target;
(2) \textbf{Misinterpretation of Signals or Obstacles}: the vehicle makes incorrect decisions in response to traffic signals or obstacles; 
(3) \textbf{Collisions}: the vehicle crashes into a static object or a dynamic road user; 
(4) \textbf{Path Planning Failures}: the vehicle cannot reach the designated goal (e.g., stops mid-route, deviates from stop points, or parks outside lane boundaries);
(5) \textbf{Smoothness Issues}: abrupt or unnatural motion (e.g., sharp turns or jerky acceleration).
Table~\ref{tab:rq2} summarizes the frequency and proportion of each type of unexpected behavior detected through scenarios generated by~\tool.
Only the last two behavior types may overlap, while others are mutually exclusive. 
Trigger conditions refer to the scenario elements most closely associated with each behavior, identified through scenario derivation analysis and Lift-based association measurement.
Through this analysis, we pinpoint the most relevant NPC information (position, type, quantity), the event or the environmental factor associated with each type of unexpected behavior.
From Table~\ref{tab:rq2}, we observe that when an NPC is positioned in front of Autoware, its path planning and decision-making modules are prone to errors. 
In addition, Autoware is more likely to experience collisions in multi-NPC scenarios, especially when oncoming vehicles interfere with its behavior. Even without collisions, sudden lane changes or turns by the NPC often cause Autoware to brake inconsistently, leading to smoothness issues.
Due to space limitations, the complete statistical data related to the triggering conditions is made publicly available.

\begin{table}[h]
\centering
\footnotesize
\caption{Occurrence, proportion, and trigger condition of unexpected behaviors}
\label{tab:rq2}
\begin{tabular}{l|ccccc}
\toprule
& \textbf{StartFail.} & \textbf{Misinterp.} & \textbf{Collisions} & \textbf{PlanFail.} & \textbf{Smooth.} \\
\hline
\textbf{Count} & 185 & 58 & 179 & 223 & 1,143 \\
\hline
\textbf{Prop.} & 10.4\% & 3.3\% & 10.2\% & 12.6\% & 64.8\% \\
\hline 
\multirow{2}{*}{\textbf{Trig.}} & \textit{R7, ah}& \textit{R7, c}& \textit{R7, km, 2}& \textit{R5, ai}& \textit{R2, ag/e}\\
\cline{2-6}
&\includegraphics[height=1em]{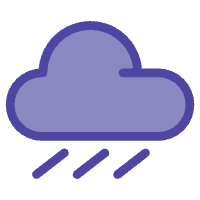}\includegraphics[height=1em]{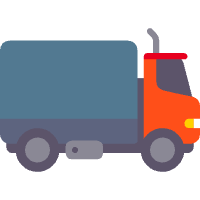} & \includegraphics[height=1em]{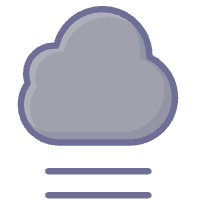}\includegraphics[height=1em]{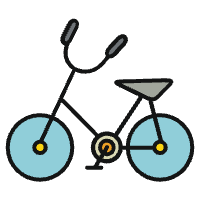} & \includegraphics[height=1em]{fog.png}\includegraphics[height=1em]{bike.png} & \includegraphics[height=1em]{rain.png}\includegraphics[height=1.2em]{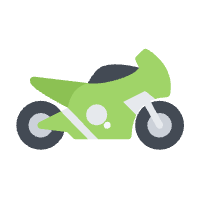}  & \includegraphics[height=1em]{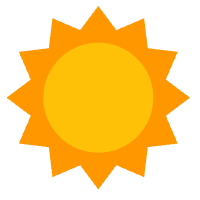}\includegraphics[height=1em]{bike.png} \\
\bottomrule
\end{tabular}
\vspace{-1.0em}
\end{table}

\begin{figure}[htbp]
    \centering
\begin{subfigure}{0.24\textwidth}
        \centering
        \includegraphics[width=0.8\linewidth]{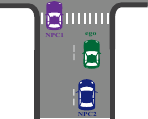}
        \caption{Blocked by NPCs}
        \label{fig:error-1}
    \end{subfigure}
    \hfill
      \begin{subfigure}{0.24\textwidth}
        \centering
        \includegraphics[width=0.8\linewidth]{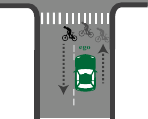}
        \caption{Cyclist Crossing}
        \label{fig:error-2}
    \end{subfigure}
    \hfill
    \begin{subfigure}{0.24\textwidth}
        \centering
        \includegraphics[width=0.8\linewidth]{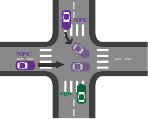}
        \caption{Intersection Crossing}
        \label{fig:error-3}
    \end{subfigure}
    \hfill
    \begin{subfigure}{0.24\textwidth}
        \centering
        \includegraphics[width=0.8\linewidth]{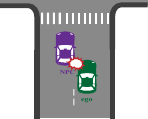}
        \caption{Rear-End Collision}
        \label{fig:error-4}
    \end{subfigure}
    
    \caption{Scenarios Leading to Ego Car unexpected Behaviors.}
    \label{fig:wrong-behavior}
    \vspace{-2.0em}
\end{figure}

Next, we analyze the unexpected behaviors of Autoware from the perspectives of scenario derivation and applied mutation operators.
For example, inserting an additional NPC and its behavior into a single-NPC scenario significantly increases the likelihood of Autoware exhibiting unexpected behaviors, with the effect particularly evident in collision scenarios—62.5\% of all collisions occur in two-NPC scenarios, even though their number is only about half that of single-NPC scenarios.
As shown in Figures~\ref{fig:error-1} and~\ref{fig:error-2}, 91.5\% of startup failures occur in scenarios where an additional NPC is placed in front of or behind Autoware, while a crossing bicycle similarly causes it to remain stationary, likely due to misinterpreting proximity as a collision risk.
As shown in Figure~\ref{fig:error-3}, assigning acceleration to the same NPC increases unexpected behaviors by up to 278 times, especially at intersections, where it causes Autoware to behave overly conservatively—stopping abruptly, refusing to proceed, or even causing collisions.
Figure~\ref{fig:error-4} shows frequent rear-end crashes, indicting that Autoware has deficiencies in dynamic object tracking and motion prediction, leading to delayed or incorrect responses.
In terms of environmental elements, Autoware performs poorly under rainy or foggy conditions, with 73.7\% of collisions occurring in such weather, indicating difficulties in handling low visibility and slippery surfaces. 
When a static obstacle is inserted—even if not blocking the path—the error rate of Autoware increases by 36.9\%, indicating misinterpretation of obstacles or inadequate path planning. 

\subsection{Bug Study}
Based on the unexpected behaviors and their trigger conditions, we analyze the behavior logic of Autoware to locate potential defects in its code or design. Below are several representative issues identified through this analysis: 
\begin{enumerate}
   \item \textit{System Crashes Due To Incorrect Path Planning.} By comparing testing results from scenarios of the same category with different initial positions of Autoware, we find that when Autoware starts in the opposite lane (with correct heading), a misalignment between the global and local paths causes it to enter an infinite loop and fail to start. 
   Source code analysis confirms that the \textit{op\_global\_planner} module generates multiple straight-line paths when multiple lane changes are detected, which leads to a mismatch between \textit{globalPathId\_roll\_outs} and \textit{globalPathId} in the \textit{op\_local\_planner}, ultimately causing a crash. This reveals a design flaw: either the number of global paths should be restricted, or the local planner should be adapted to handle ID mismatches more robustly.


    \begin{lstlisting}[language=C++, caption={Autoware trajectory synchronization logic}]
    // globalPathId_roll_outs is the gid of the first Waypoint object in the last Lane object from the /local_trajectories message.
    // globalPathId is the gid of the first Waypoint object in the first Lane object from the /lane_waypoints_array message. 
    if(globalPathId_roll_outs == globalPathId) 
    {
        bWayGlobalPath = false; // If this variable is not true, the program will keep looping
        m_GlobalPathsToUse = m_GlobalPaths;
        std::cout << "Synchronization At Trajectory Evaluator: GlobalID: " <<  globalPathId << ", LocalID: " << globalPathId_roll_outs << std::endl;
    }
    \end{lstlisting}
    
    \item \textit{Node Startup Order Causing Launch Failures.} By comparing scenarios derived from a single-NPC setup to more complex multi-NPC variants, we observe that Autoware occasionally fails to start, even when a valid path is available. This issue arises more frequently in derived scenarios with increased complexity, where longer initialization times cause the destination to be published after the control modules (e.g., \textit{pure\_pursuit}, \textit{twist\_filter}) have already started. In such cases, the control modules enter an unrecoverable waiting state due to the absence of mechanisms for monitoring and recovering late-arriving path inputs. This reflects a coordination flaw between planning and control components in Autoware. To ensure the smooth execution of our experiments, we patch the launch files to enforce a proper node startup sequence.

    \item \textit{Startup Failure Under Low-Adhesion Conditions}. In RQ2, we observe that Autoware occasionally fails to start under rainy and slippery road conditions. Despite the continuous issuance of control instructions, Autoware remains stationary. Reproduction experiments show that this issue is caused by the control module applying both high throttle and a large steering angle during the initial acceleration phase, leading to front-wheel slippage and a lack of effective traction. This combination of control inputs violates the physical constraints on tire forces and thus cannot be properly executed by the simulation engine. This indicates that the current control strategy lacks proper feasibility checks under low-adhesion conditions.

\end{enumerate}

\subsection{Threats to validity}
\subsubsection{The threats to internal validity.}
The threats to the internal validity of \tool~mainly come from the quality of seed scenario files and the accuracy of ground-truth construction. 
After generating OpenSCENARIO files from accident reports, we perform both syntactic and semantic validation. 
We use \textit{xmllint} to ensure conformance with the XSD schema and simulate each scenario to verify that key elements—such as entity types, positions, and event sequences—align with the original report. 
Each seed scenario is independently reviewed by three authors using a predefined checklist, and disagreements are resolved through inter-rater agreement analysis and group discussion. 
As the scenario information is extracted by an LLM, occasional misinterpretations may arise due to textual ambiguity. These are manually inspected and corrected before scenario generation.

\subsubsection{The threats to external validity.}
The main threat to external validity lies in the generalizability of generated scenarios across simulators and report sources. 
To address platform differences, \tool~generates OpenSCENARIO-compliant files and parses OpenDRIVE maps to build a reusable road network database, enabling execution across multiple simulators. 
To accommodate variability in textual reports, \tool~automatically checks whether required behavioral elements—such as vehicle actions, relative positions, and event sequences—are present. If any critical information is missing, \tool~flags the absence and prevents scenario generation. 
This design ensures that only complete and valid inputs are used, improving the portability and robustness of generated scenarios. 
Future work will focus on supporting additional report formats and scenario types to further broaden applicability.

\section{Related Work}

Test scenario generation based on traffic accident reports aims to extract key contextual information and convert it into essential simulation components to construct realistic scenarios for ADS testing. Accident reports exist in multiple forms, each requiring different extraction techniques. For example, the M-CPS model~\cite{zhang2023building} processes images and videos by segmenting traffic participants, enabling scene reconstruction. However, such resources are often limited by privacy regulations and data availability, and their processing requires high-performance hardware and substantial storage space.

To address these issues, recent methods focus on textual reports. Representative approaches include ADEPT~\cite{wang2022adept}, SoVAR~\cite{guo2024sovar}, and LeGEND~\cite{tang2024legend}. ADEPT extracts information from text and generates adversarial scenarios by optimizing Scenic templates using the feedback of ADSs, but incurs high computational costs and suffers from simulation-to-reality gaps. Both SoVAR and LeGEND generate perception-layer inputs by extracting scenario information from textual accident reports and directly invoking the APIs of the LGSVL simulator—SoVAR based on key event extraction, and LeGEND via a domain-specific language (DSL) that maps text to logical scenarios.
However, these methods face several limitations: (1) the LGSVL simulator has not been actively maintained since 2022~\cite{lgsvl}, making it incompatible with modern ADSs and their evolving interface requirements; (2) their heavy reliance on simulator-specific APIs results in strong platform coupling, making it difficult to migrate their testing pipelines to other widely used or better-supported simulators such as CARLA or Autoware-compatible environments;(3) by directly generating perception-level data instead of standardized scenario files, both methods lack reusable and portable outputs such as OpenSCENARIO files, reducing their value in system-level testing, comparative analysis, and reproducibility.
In contrast, \tool~produces standardized, extensible scenario files in the OpenSCENARIO format, supports simulator-agnostic workflows, and enables cross-platform, scalable ADS testing.

\section{conclusion}

In this paper, we propose \tool, a method for generating OpenSCENARIO files from textual accident reports. \tool~employs the LLM to convert reports into seed scenarios, which it expands through disassembly, mutation, and assembly to improve scenario diversity.
In the experiments, \tool~generates 4,373 valid scenarios from 33 seed scenarios, and then employs these generated scenarios to test Autoware.
The experimental results show that the scenarios generated by \tool~are syntactically and semantically valid, highly diverse, and effectively uncover various unexpected behaviors of Autoware in terms of safety, smoothness, and smartness. 
By analyzing these behaviors and mapping them to the source code of Autoware, we identify specific design and implementation defects.
These findings show that \tool~can help developers understand system behavior, diagnose issues, and improve the performance of ADSs. 

\section*{Data Availability}

The source code of \tool, the seed generation files converted by \tool, the new scenario files generated by \tool, and the testing results related to the unexpected behaviors of Autoware are publicly available on our project website.

\balance
\bibliographystyle{IEEEtran}
\bibliography{ref}
\end{document}